# STUDYING STRUCTURE FORMATION WITH LARGE GALAXY REDSHIFT SURVEYS[†]

DAVID H. WEINBERG
*Institute for Advanced Study*
*Princeton, NJ 08540, USA*
dhw@guinness.ias.edu

## ABSTRACT

I outline the connections between some of the most widely used statistical measures of galaxy clustering and the fundamental issues in the theory of structure formation. I devote particular attention to the problem of biasing, *i.e.* to a possible difference between the distribution of galaxies and the distribution of mass. Using numerical experiments, I show that a local morphology-density relation leads to different slopes and amplitudes for the correlation functions of different galaxy types on small scales, but that the correlation functions on large scales all have the same shape, differing only by constant factors. I also examine a variety of biasing models in which the efficiency of galaxy formation depends on local properties of the mass distribution. While these purely local influences do lead to a large-scale bias, in all cases the bias factor becomes constant on scales larger than the galaxy correlation length.

## 1. Introduction

Galaxy redshift surveys are driven in part by the map-maker's instinct, by an urge to discover, name, classify, and measure the structures that we can see through our telescopes. Redshift surveys are also driven by a more theoretical objective: they provide the data with which we can test our ideas about the origin of structure in the universe. There are various ways to state the "big questions" in the field of structure formation, but any complete list would include the following. Did structure form by gravitational instability, as assumed in most of the leading theories? If so, what were the properties of the primordial fluctuations that seeded the growth of structure, and what physical process created them? What is the dark matter? What are the values of the density parameter, $\Omega$, and the cosmological constant, $\Lambda$? The answers to these questions are intimately linked to theories of particle physics and to theories of the origin, early history, and ultimate fate of the universe.

In addition to these "fundamental" questions, there is a second category of important, "astrophysical" questions, related to galaxy formation. How did galaxies form? What physical processes, besides gravity, played an important role? What processes determine galaxy luminosity, size, color, and morphology? What is the relation between the distribution of galaxies and the underlying distribution of mass? Because galaxies are the markers by which we trace large-scale structure, we cannot address questions in the first category without simultaneously addressing those in the second, especially the final question about the relation between galaxies and mass.

This talk has two main parts. In the first, I will sketch the connections between statistical measures of galaxy clustering and the questions listed above. In the second, I will focus more specifically on the most vexing problem that afflicts the interpretation of large-scale structure data, the uncertain relation between galaxies and mass. Before either of these discussions, I must define a few basic terms.

## 2. Some Definitions

It is convenient to describe fluctuations of the density field in terms of the dimensionless density contrast,

$$\delta(\mathbf{x}) \equiv \frac{\rho(\mathbf{x}) - \bar{\rho}}{\bar{\rho}} , \qquad (1)$$

or its Fourier transform,

$$\tilde{\delta}(\mathbf{k}) \equiv \int_V d^3 x \delta(\mathbf{x}) e^{i\mathbf{k}\cdot\mathbf{x}} . \qquad (2)$$

Important statistical descriptors of these fields are the correlation function,

$$\xi(r) \equiv \langle \delta(\mathbf{x})\delta(\mathbf{x}+\mathbf{r})\rangle, \qquad (3)$$

and its Fourier transform, the power spectrum,

$$P(k) \equiv \langle \tilde{\delta}(\mathbf{k})\tilde{\delta}^*(\mathbf{k})\rangle/V, \qquad (4)$$

where $V$ is the integration volume in Eq. (2). If the complex phases of the Fourier modes $\tilde{\delta}(\mathbf{k})$ are randomly distributed, then $\delta(\mathbf{r})$ is a Gaussian random field, and the power spectrum completely specifies its statistical properties.

When $\delta \ll 1$, linear perturbation theory implies that

$$\delta(\mathbf{x}, t_2) = \delta(\mathbf{x}, t_1) D(t_2)/D(t_1), \qquad \tilde{\delta}(\mathbf{k}, t_2) = \tilde{\delta}(\mathbf{k}, t_1) D(t_2)/D(t_1) . \qquad (5)$$

Since Eq. (5) gives a trivial recipe for scaling linear fluctuations from one epoch to another, one can usually afford to play a a bit loose with the phrase "initial conditions." The linear growth factor $D(t)$ depends on the values of $\Omega$ and $\Lambda$. For $\Omega = 1$, $\Lambda = 0$, the behavior is especially simple: $D(t) \propto t^{2/3} \propto (1+z)^{-1}$.

If galaxies form with different efficiency in different environments, then the galaxy distribution may paint a biased picture of the underlying mass distribution. The simplest model that describes a difference between the galaxy density contrast and the mass density contrast, both smoothed over some scale $R_s$, is the linear bias model,

$$\delta_g(\mathbf{x}, R_s) = b\delta(\mathbf{x}, R_s) , \qquad (6)$$

where $b$ is the bias factor. When the galaxy-mass relation is more complicated than this, different definitions of the bias factor can yield different values. One often defines $b$ to be the ratio of the rms galaxy fluctuation to the rms mass fluctuation on some scale. Related definitions include $b^2 = \xi_{gg}(r)/\xi(r)$ and $b^2 = P_g(k)/P(k)$. Bias factors derived from comparisons of galaxy density and velocity fields usually involve some overall fit to Eq. (6). In principle the bias factor can vary with scale; I return to this issue in §4.

The linear bias model must break down when $b > 1$ and $\delta < -1/b$, since it predicts negative galaxy densities. A more general and less problematic assumption is that of a local relation between the galaxy and mass density fields,

$$\delta_g(\mathbf{x}, R_s) = f[\delta(\mathbf{x}, R_s)] , \qquad (7)$$

where $f$ is an arbitrary function. When the fluctuation amplitude is small, it is useful to consider a Taylor expansion of this relation,

$$\begin{aligned} \delta_g &= a + b_1 \delta + \frac{b_2 \delta^2}{2} + \frac{b_3 \delta^3}{6} + \dots , \\ b_1 &= f'(0), \quad b_2 = f''(0), \quad b_3 = f'''(0), \quad \dots . \end{aligned} \qquad (8)$$

The constant $a$ is fixed at any order by the requirement $\langle \delta_g \rangle \equiv 0$. The linear bias model (6) emerges as the first-order expansion of this local model, with $b = b_1$, suggesting that linear bias may often be an adequate description on scales where linear perturbation theory applies. More generally, the local model is characterized by the hierarchy of bias factors $b_1$, $b_2$, $b_3$, .... Once again, these bias factors can in principle depend on scale. Note that one can easily construct models in which rms galaxy and mass fluctuations are equal but galaxies still do not trace mass in detail.

## 3. Clustering Statistics and the Physics of Structure Formation

I will now sketch connections between some of the most widely used statistical measures of galaxy clustering and the questions raised in the Introduction. I will concentrate on large-scale measures, where the connections are most direct and where the new generation of redshift surveys will make the largest incremental change to our knowledge. I will skip over many subtleties. A more detailed discussion of clustering statistics, focusing on somewhat smaller scales, appears in reference 1.

*3.1. Power Spectrum and Autocorrelation Function*

Since the power spectrum and the correlation function are a Fourier transform pair, perfect knowledge of one affords perfect knowledge of the other. However, estimates of these quantities from noisy data need not be equivalent, so in practice it is desirable to measure them independently. Theoretical discussions are usually couched in terms of the power spectrum, and conventional wisdom holds that $P(k)$ can be measured more accurately than $\xi(r)$ at large scales. Michael Vogeley's contribution to this volume discusses estimation of the power spectrum, results from recent redshift surveys, and prospects for future measurements.

The simplest versions of inflation and of topological defect theories (strings, textures, etc.) produce fluctuations that are scale-invariant — they enter the horizon with an rms amplitude that is independent of time. For perturbations that enter the horizon after the epochs of matter-radiation equality and recombination, scale-invariance leads to the Harrison-Peebles-Zel'dovich power spectrum, $P(k) \propto k$. On smaller scales the power spectrum shape is $P(k) \propto k T^2(k)$, where the transfer function $T(k)$ describes the growth history of a perturbation of wavenumber $k$ between the time it enters the horizon and the time that the universe becomes matter-dominated and effectively pressureless. The transfer function depends on the cosmic equation of state during the transition from radiation to matter domination, and the power spectrum therefore encodes information about the material content of the universe, in particular about the nature and amount of dark matter. In inflationary cold dark matter (CDM) models, the linear-theory power spectrum turns over at the scale representing the horizon at matter-radiation equality, and it gradually steepens to $k^{-3}$ on very small scales. The physical value of this turnover scale, measured in $h^{-1}$Mpc, depends on the parameter combination $\Omega h$.

High precision measurements of the power spectrum near the turnover, which should be possible with redshift surveys like 2dF and Sloan, are the key to obtaining good physical constraints on the contents of the universe. It would be especially exciting to detect the subtle oscillations that arise from photon-baryon acoustic waves, as these would provide an impressive confirmation of our theoretical picture, a measurement of $\Omega_{baryon}$, and a way to distinguish between adiabatic fluctuations (predicted by inflation) and isocurvature fluctuations (predicted by topological defect models). Departures from scale-invariance can also affect the shape of the power spectrum, and $P(k)$ measurements alone cannot distinguish such departures from transfer function effects. Independent constraints, especially from microwave background measurements, can help achieve this separation. Clear departures from scale-invariance would provide vital clues about the early-universe physics that generates fluctuations.

Current theories for the origin of fluctuations do not predict their magnitude *a priori*, so the amplitude of the power spectrum provides a normalization constraint, but it does not tell us much about the physical source of structure. In the linear bias model the mass power spectrum is just $1/b^2$ times the galaxy power spectrum. On small scales, the power spectrum and the correlation function are affected by non-linear dynamics and by the details of biasing. They still provide observational constraints that a complete theory must satisfy, but their relation to the basic questions of structure formation is complicated and indirect.

*3.2. Anisotropy of Clustering in Redshift Space*

I have thus far ignored the fact that we observe galaxies in redshift space rather than real space. Peculiar velocities distort clustering in redshift space by changing the apparent distances of galaxies, turning the line of sight into a preferred direction. One can recover real-space values for the power spectrum and correlation function by projection and inversion. However, the distortions themselves contain important information. By comparing clustering along the line of sight to clustering across the line of sight, one can infer statistical information about galaxy peculiar velocities, without measuring the motion of any individual galaxy.

Kaiser[2] derived the anisotropy of the power spectrum induced by peculiar motions in the linear regime, where inflows to high-density regions and outflows from low-density regions enhance clustering along the line of sight. In the absence of bias, a measurement of this anisotropy yields the value of $\Omega$. With linear bias, the anisotropy depends on the parameter combination $\beta \equiv \Omega^{0.6}/b$. There are three broad approaches to measuring this anisotropy in real redshift surveys, one based on the correlation function,[3,4] one on the power spectrum,[5] and the third on spherical harmonics.[6,7] Each has its advantages and disadvantages, and it remains to be seen which method will be the most powerful when applied to the next generation of redshift surveys.

Virial velocities in collapsed objects create "fingers of God," and these cause redshift-space anisotropies that are opposite in sign from those produced by large-scale, coherent flows. If one wants to measure the linear-theory distortions, then the virial velocities are an enormous nuisance, demanding complicated and uncertain corrections. However, one can use clustering anisotropy on small scales to quantify the virial velocities themselves, the classic approach in this regard being the measurement of the galaxy pairwise velocity dispersion.[4,8] The pairwise dispersion depends on $\Omega$ and on the amplitude of mass fluctuations, and for this reason it is often invoked as a cosmological test. However, as I will argue in §4, it is also sensitive to the intricate details of biasing, and this sensitivity limits its usefulness as a constraint on theoretical models.

*3.3. The Probability Distribution Function*

From the correlation function or the power spectrum, one can compute the rms fluctuation of the galaxy density field smoothed on any scale, but rms fluctuations do not tell the whole statistical story. At any smoothing, one can examine the full probability distribution function (PDF) of the density field, *i.e.* the probability $P(\delta_g)d\delta_g$ that the smoothed galaxy density contrast at a randomly chosen position has a value between $\delta_g$ and $\delta_g + d\delta_g$. Apart from technicalities related to discreteness corrections, the PDF is equivalent to the count-in-cell distributions $f(N)$,[9] which can themselves be related to the higher-order correlation functions.[10] The PDF encodes similar information to the higher-order correlations, in less detailed but more digestible form.

The galaxy PDF is of course related to the PDF of the initial fluctuations (the IPDF). Standard inflationary models predict a Gaussian IPDF, while topological defect models and some specialized versions of inflation produce non-Gaussian initial conditions. The central limit theorem ensures that Gaussian fluctuations are somewhat generic (even the defect models are not strongly non-Gaussian), so evidence for a Gaussian IPDF provides a reassuring indication that our picture of structure formation is on the right track, but only moderate support for an inflationary origin of perturbations. Convincing evidence for a non-Gaussian IPDF would be quite exciting, since the departure from Gaussianity would be a specific clue to the physical origin of fluctuations.

The galaxy PDF is affected by non-linear gravitational evolution and by by biasing. (It also suffers from redshift-space distortions, but these can be removed by projection.) One can compute theoretical predictions using numerical simulations, since they automatically include non-linear effects and can incorporate models of biased galaxy formation. However, on large scales one can make great progress analytically, treating non-linear evolution via perturbation theory. In the case of Gaussian initial conditions, perturbation theory yields remarkably simple results for the moments of the PDF. For example, with a top-hat smoothing filter and a power-law power spectrum $P(k) \propto k^n$, the ratio of the third moment to the square of the variance is

$$\frac{\langle\delta^3\rangle}{\langle\delta^2\rangle^2} \equiv S_3 = \frac{34}{7} - (3+n) , \qquad (9)$$

independent of the smoothing scale itself.[11] Computation of the skewness $\langle\delta^3\rangle$ requires second-order perturbation theory, so it is risky to apply a linear bias model even at large scales, but one can use the second-order Taylor expansion of Eq. (8) to show that a non-

linear, local bias preserves the form of Eq. (9), while changing the constant to[12,13]

$$S_{3g} \equiv \frac{\langle \delta_g^3 \rangle}{\langle \delta_g^2 \rangle^2} = \frac{S_3}{b_1} + \frac{3b_2}{b_1^2} \ . \tag{10}$$

Continuing to higher orders, perturbation theory yields a sequence of hierarchical relations between moments of the density field and the variance.[14] When the fluctuation amplitude is small, one can plug these moment relations into the Edgeworth series to obtain the non-linear PDF.[13,15] Large-scale measurements of the PDF thus allow one to test the combined hypothesis of gravitational instability, Gaussian initial conditions, and local biasing.

Topological statistics like the genus[16] or crossing-frequency[17] of isodensity contours provide an entirely independent way to test the Gaussian hypothesis. The volume-weighting approach advocated by Gott and collaborators removes any dependence on the PDF, so that these statistics examine the global organization of fluctuations rather than the distribution of their amplitudes.

*3.4. Properties of Galaxy Clusters*

Rich galaxy clusters are the most massive virialized (or at least nearly virialized) systems in the universe, and in standard scenarios they form by the gravitational collapse of volumes $5 - 10h^{-1}$Mpc in radius. The most basic statistic that characterizes the population of galaxy clusters is the mass function $n(M)$, where $n(M)dM$ is the number of clusters per unit volume with mass in the range $M \to M + dM$. The mass function depends on the fluctuation amplitude at $R \sim 5 - 10h^{-1}$Mpc, which determines the probability that such volumes collapse, and on the value of $\Omega$, which determines the mass contained in such volumes. The dependence is quite sensitive because clusters form from rare perturbations on the tail of the initial probability distribution, and the cluster mass function therefore yields a powerful constraint despite the ambiguities associated with cluster identification and mass estimation. White *et al.*[18] conclude that

$$\sigma_8 \Omega^{0.56} = 0.57 \tag{11}$$

for a broad class of theoretical models, where $\sigma_8$ is the rms mass fluctuation in $8h^{-1}$Mpc spheres. This constraint is probably accurate to 15%, unless our understanding of cluster masses is seriously amiss (and the weak lensing results discussed by Kaiser in this volume suggest that it might be).

The attractive feature of Eq. (11) is that it constrains the *mass* fluctuation amplitude, while measurements of the correlation function or the power spectrum constrain the galaxy fluctuation amplitude. Comparing the two yields a relation between $\Omega$ and the bias factor at the $8h^{-1}$Mpc scale. The correlation function of the CfA1 survey, for instance, yields $\sigma_{8,g} \approx 1$,[8] implying

$$b_8 \equiv \frac{\sigma_{8,g}}{\sigma_8} \approx 1.75\Omega^{0.6}. \tag{12}$$

By comparing cluster mass-to-light ratios to the global mass-to-light ratio that is required to close the universe, one can obtain an estimate of $\Omega/b_c$. Here $b_c$ is a "cluster bias factor," the ratio of the light overdensity in clusters to the mass overdensity in clusters. For typical mass-to-light ratios $\sim 300h$, this exercise yields $\Omega/b_c \sim 0.2$. It is important to remember, however, that $b_c$ could be quite different from bias factors defined in terms of rms statistics, since the linear bias model is sure to break down at the high density contrasts of galaxy clusters.

If the universe is open, present-day clusters should have assembled most of their mass at moderate redshifts, before curvature came to dominate the cosmic expansion. If $\Omega = 1$, on the other hand, the outer regions of clusters should have collapsed quite recently. The evolution of the cluster population thus offers a different route to constraining $\Omega$, via the history of clustering. This approach is tricky to implement observationally, since one can easily make the mistake of comparing high-redshift apples to low-redshift oranges. As an alternative, one can hope that archaeology of present-day clusters will reveal the history of their formation, with clues embedded in density profiles[19,20] or in dynamically young substructure.[21]

*3.5. Clustering of Different Galaxy Types*

Galaxies come in a wide range of shapes and sizes. In some cases we have clear evidence that different types of galaxies cluster differently (*e.g.* ellipticals vs. spirals), while in other cases (*e.g.* bright galaxies vs. faint galaxies) the current evidence is more ambiguous. New surveys of larger samples should improve our knowledge in this area, especially the Sloan survey, which will combine excellent photometric data with a redshift sample large enough to allow accurate clustering measurements for finely divided sub-classes. Studies of the clustering of different galaxy types will play an important role in our attempts to answer the "astrophysical" questions listed in the Introduction. If we want to know what processes influence galaxy luminosity, for example, it helps to know as much as we can about the environments in which luminous and faint galaxies form. Of course such information can admit a variety of interpretations, witness the long-standing nature vs. nurture debate over the origin of ellipticals.

By comparing the clustering of different galaxy types, we also gain some clues about the relation between galaxies and mass. Thus far these clues are largely negative ones, *e.g.* the difference in the clustering amplitudes of optical and IRAS galaxies tells us that they cannot both be unbiased tracers of the mass distribution. Higher-precision studies that extend to larger scales could yield important positive results. For instance, if we find the same shape for the power spectra of bright galaxies and faint galaxies, ellipticals and spirals, emission-line galaxies and quiescent galaxies, then it would be hard to imagine that this is not also the shape of the mass power spectrum. If different galaxy types yield incompatible results even at large scales, then we will at least know that we should be cautious in drawing conclusions about the mass distribution.

## 4. Local Galaxy Formation

A constant thread running through the above discussion is the uncertainty associated with biased galaxy formation. The blame for this uncertainty lies with the theorists rather than the observers; if we were doing our jobs properly, then we would be able to go from specified primordial fluctuations directly to the predicted distribution of galaxies. In the long run, cosmological simulations with gas dynamics should provide just such *ab initio* predictions of galaxy clustering, but at present these simulations suffer from numerical limitations and poor statistics, because even the most powerful supercomputers can barely achieve the dynamic range that such studies demand. As a complementary approach, I have investigated the effects of applying a variety of simple, *ad hoc* biasing schemes to mass distributions from cosmological N-body simulations. The common feature of the schemes that I consider is that they are local: when a particle in the simulation decides whether to become a "galaxy," it can ask questions only about the mass distribution in the nearest few Mpc. These questions can be rather general, so the definition of local bias used here is broader than the local galaxy-mass relation of Eq. (7).

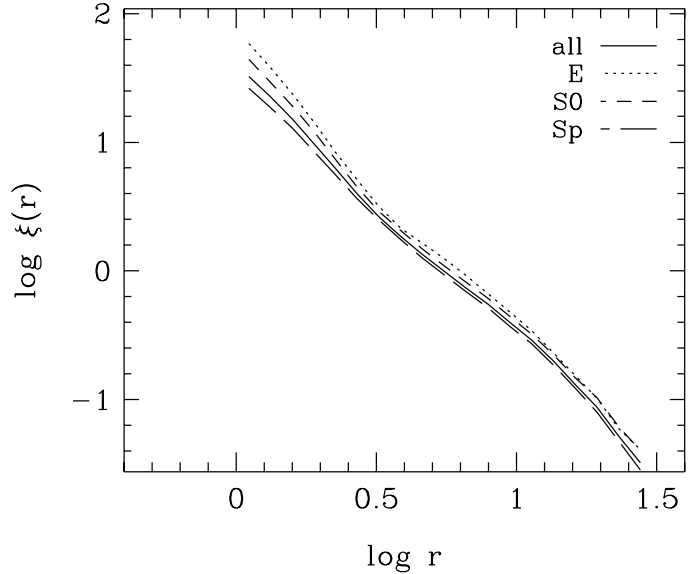

Figure 1: Correlation functions for all galaxies (solid line), ellipticals (dotted line), S0's (short-dashed line), and spirals (long-dashed line) in a low-density CDM model, with $r$ in $h^{-1}$Mpc.

I begin with an example that may be of interest even to skeptics, who think that the term "biased" applies to theorists rather than to galaxies. It is well known that early-type galaxies reside preferentially in high-density environments, while field galaxies are most likely to be late-type spirals. What effect does this morphological segregation have on the clustering of different galaxy types, as measured by the correlation function?

The solid line in Figure 1 shows $\xi(r)$ for the full galaxy distribution in simulations of a low-density, spatially flat, cold dark matter universe ($\Omega = 0.3$, $\Lambda/3H_0^2 = 0.7$, $\Omega h = 0.25$). I assume that the galaxy population as a whole traces the mass, so this line is also the mass correlation function. To divide the sample into morphological classes, I estimate the local density around each galaxy from the distance to the fourth-nearest neighbor, then assign the galaxy a morphological type following a piecewise power-law fit to the morphology-density relation of the CfA redshift survey.[22]

Dotted, short-dashed, and long-dashed lines in Figure 1 show the correlation functions of the elliptical, S0, and spiral galaxies identified in this manner. At small scales ($r < 3h^{-1}$ Mpc), the correlation functions of the early-type galaxies are both higher and steeper than that of the full galaxy population, while the spiral correlation function is slightly lower and flatter. This behavior accords well with observed angular correlations.[23,24] Observational constraints at large separations are presently rather poor, and if one were simply to fit power laws to the small-scale correlations and extrapolate, one would expect the correlation functions to cross, with spirals more strongly correlated than ellipticals at large separations. However, the model in Figure 1 does not behave this way at all: at $r > 3h^{-1}$ Mpc the correlation functions all acquire the *same* shape, and they *never* cross. Indeed, the early-type galaxies show stronger clustering than the spirals even at $r = 20h^{-1}$ Mpc, by about 30% in $\xi(r)$. The value of this large-scale bias is sensitive to the adopted parameters of the morphology-density relation, and the simulations (which use a particle-mesh code with $0.5h^{-1}$ Mpc cells) certainly underestimate the effect, because they underestimate the densities in the cores of collapsed clusters and groups.

If morphological segregation in the real universe is adequately described by a local morphology-density relation, then the qualitative behavior of the large-scale correlation functions should resemble that in Figure 1. Note that purely *local* segregation does give rise to a bias at large scales. The reason for this large-scale amplification is explained in Kaiser's classic paper: in a model with Gaussian initial conditions, rich clusters form preferentially in regions of high background density.[25] One can think of the ellipticals and S0's as inheriting a share of the higher-amplitude, cluster correlation function.

For a more detailed exploration of biasing effects, I turn to simulations of the standard CDM model ($\Omega = 1$, $\Omega h = 0.5$), normalized so that $\sigma_8 = 0.5$. I use five different prescriptions to select biased subsets of the particle distribution. In all five schemes, the probability that a particle is labeled as a "galaxy" depends only on properties of the mass distribution in a sphere of radius $4h^{-1}$Mpc around it. In the simplest scheme, the bias is a step-function in density: all particles below some threshold density are eliminated, and all particles above the threshold have an equal probability of becoming a galaxy. The second scheme uses a threshold in the local ram pressure ($\rho\sigma^2$, where $\sigma$ is the velocity dispersion within the $4h^{-1}$Mpc sphere). I also consider a non-linear density bias, in which the relation between galaxy and mass densities is roughly a power law. The last two schemes combine a density threshold with a geometrical anisotropy criterion, based on the eigenvalues ($\lambda_1 > \lambda_2 > \lambda_3$) of the local moment-of-inertia tensor. "Sheets biasing" selects particles with large values of $\lambda_1/\lambda_3$, and "filaments biasing" selects particles with large values of $\lambda_1/\lambda_2$. The motivation for these two schemes is simply that many galaxies are observed to lie in thin sheets and filaments, and this tendency could reflect a link between the geometry of large-scale collapse and the efficiency of galaxy formation. However, none of these schemes has a compelling physical argument behind it. I am using them as a way to parametrize our ignorance about galaxy formation, and I hope that they are diverse enough to encompass the behavior of most local biasing models.

Each of these prescriptions has a single free parameter whose value controls the strength of the bias, *e.g.* the value of the threshold in the simple density scheme. In each case, I choose this parameter so that the rms fluctuation of the galaxy density field is twice that of the mass density field when both are smoothed with a Gaussian filter of radius $10h^{-1}$Mpc. On this scale, therefore, the conventionally defined bias factor is $b = 2$.

The solid line in Figure 2 shows the mass correlation function of these simulations, multiplied by a factor of four. Other lines show correlation functions of the galaxy populations selected by the various biasing schemes. On small scales ($r \lesssim 4h^{-1}$Mpc), different schemes yield correlation functions of different amplitude and different slope, with the sharp-threshold prescriptions yielding the lowest and shallowest functions and the geometrical prescriptions yielding the highest and steepest. In this regime, the bias factor defined by $b^2 = \xi_{gg}(r)/\xi_{\rho\rho}(r)$ varies with $r$, in a way that depends on the adopted biasing prescription. However, at large separations all five correlation functions have the same shape as the mass correlation func-tion, and $b$ is independent of scale. There is a small offset in the large-scale amplitude of $\xi(r)$ for the geometrical models, which arises from the normalization at $10h^{-1}$Mpc. The rms fluctuation is given by an integral over the correlation function, so the models with the strongest small-scale correlations must have weaker large-scale correlations in order to satisfy the normalization constraint.

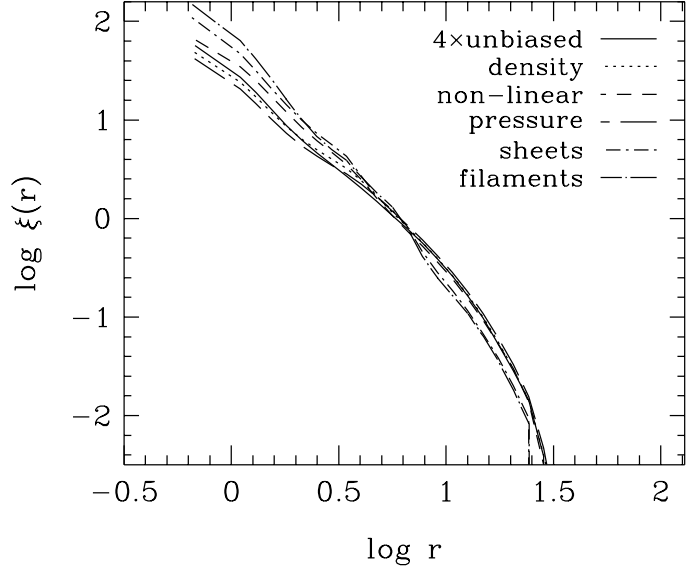

Figure 2: Correlation functions in the standard CDM model for different biasing schemes.

Results for the power spectrum are similar to those for the correlation function. At small scales (high $k$) the different biasing models yield power spectra of different amplitudes and shapes, but at long wavelengths the galaxy power spectra are scale-independent multiples of the underlying mass power spectrum, with a small amplitude offset between the geometrical models and the others.

Babul & White[26] and Bower et al.[27] have proposed biasing schemes that lead to a scale-dependent bias factor even at large separations, and they argue that such a scheme could reconcile the standard CDM model with observational evidence that favors a redder power spectrum. Both schemes share the property that they are non-local, i.e. the efficiency of galaxy formation depends directly on events (quasar formation, for example) happening $10-20h^{-1}$Mpc away. The results in Figure 2 suggest that non-locality is an essential rather than an incidental feature of these models, and that any local model will yield a constant bias factor on scales that are large compared to (a) the galaxy correlation length and (b) the scale over which the environment influences the efficiency of galaxy formation. Coles[28] presents an analytic argument that supports this point of view.

What about higher-order correlations? Investigation of skewness suggests that, at least for this statistic, these bias models behave much like the local density model of Eq. (7). The ratio $S_{3g} \equiv \langle \delta_g^3 \rangle / \langle \delta_g^2 \rangle^2$ is approximately independent of smoothing scale, but the value of $S_{3g}$ depends on the biasing model, ranging from $S_{3g} \approx 1.5$ for pressure bias to $S_{3g} \approx 3.5$ for filament bias. In the language of Eq. (8), these schemes have the same $b_1$ but different values of $b_2$.

Figure 3 shows the pairwise velocity dispersion for the mass distribution (solid line) and the galaxy distributions. Even though biasing does not alter any particle velocities, $\sigma_v(r)$ depends strongly on the adopted biasing scheme, especially at separations $r < 2h^{-1}\text{Mpc}$, where it varies by nearly a factor of two. Because contributions to $\sigma_v$ are weighted by the number of pairs and by the square of the velocity difference, the statistic is sensitive to the representation of galaxies in rich clusters. The non-linear density bias puts the largest fraction of galaxies in clusters, so it gives the highest pairwise dispersion. The geometrical schemes give the lowest dispersions on small scales, even though they have the highest spatial correlations. The pairwise dispersion is not a reliable statistic with which to test theories of structure formation because the predicted values are sensitive to the most uncertain aspect of the theoretical models, the relation between galaxies and mass in dense regions. The cluster mass function discussed in §3.4 offers a more robust way to get at similar information, since in this measure the weight of a cluster does not depend on the number of galaxies it contains (provided it has enough galaxies to be identified as a cluster in the first place).

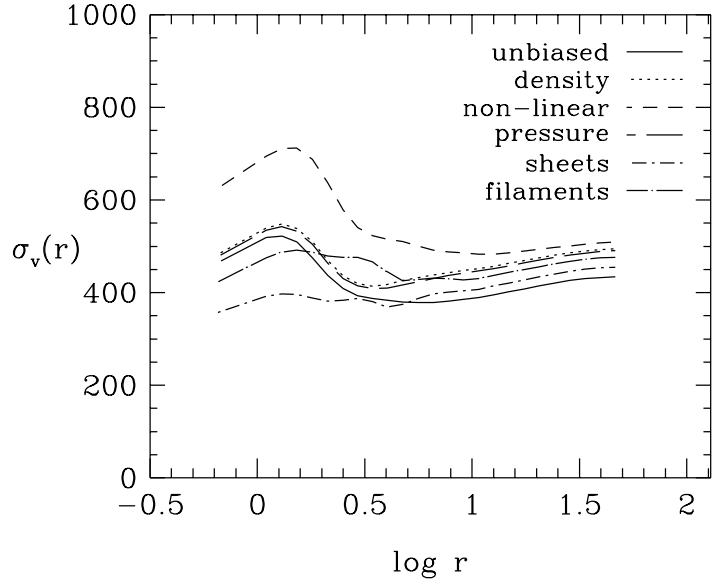

Figure 3: Pairwise velocity dispersions (in km/s) for the standard CDM model, with different biasing schemes.

## 5. Conclusions

The next generation of galaxy redshift surveys will allow us to make high-precision measurements of large-scale structure over an enormous dynamic range. This precision, especially on scales that are still close to the linear regime, is the key to answering the physical questions raised in the Introduction. The connections between these questions and the statistics of galaxy clustering are complex, but there are enough overlapping constraints and distinguishing tests that analysis of large, uniform data sets should bring us much closer to understanding how structure in the universe formed.

The local galaxy formation experiments discussed in §4 are somewhat encouraging. The assumption that the efficiency of galaxy formation depends on local properties of the mass distribution seems to remove much of one's freedom to fit arbitrary theories to observational data. In particular, the models examined all lead to a scale-independent bias on large scales, and they preserve a hierarchical relation between the skewness and the variance of the galaxy distribution. The assumption of local galaxy formation is physically plausible, and it is general enough to encompass many of the ideas that theorists have proposed for biasing the galaxy distribution, but nature can always be inventive. Large-scale clustering studies of different galaxy types will allow us to distinguish a locally biased universe from a more complicated, but perhaps even more interesting universe in which forming galaxies pay direct attention to distant events.

I thank the organizers for putting together an exciting workshop, Changbom Park for the use of his N-body code, and colleagues too numerous to mention for stimulating discussions on the topics covered in this talk. I acknowledge financial support from the W. M. Keck Foundation and NSF grant PHY92-45317.